# The Algorithmic State Architecture (ASA): An Integrated Framework for AI-Enabled Government

Zeynep Engin*[1,2], Jon Crowcroft[3,4,1], David Hand[5,1], Philip Treleaven[2]

[1] Data for Policy CIC, London, United Kingdom
[2] Department of Computer Science, University College London (UCL), United Kingdom
[3] Department of Computer Science and Technology, University of Cambridge, United Kingdom
[4] The Alan Turing Institute, London, United Kingdom
[5] Department of Mathematics, Imperial College London, United Kingdom

* Corresponding Author: z.engin@dataforpolicy.org

## Abstract

As artificial intelligence transforms public sector operations, governments struggle to integrate technological innovations into coherent systems for effective service delivery. This paper introduces the Algorithmic State Architecture (ASA), a novel four-layer framework conceptualising how Digital Public Infrastructure, Data-for-Policy, Algorithmic Government/Governance, and GovTech interact as an integrated system in AI-enabled states. Unlike approaches that treat these as parallel developments, ASA positions them as interdependent layers with specific enabling relationships and feedback mechanisms. Through comparative analysis of implementations in Estonia, Singapore, India, and the UK, we demonstrate how foundational digital infrastructure enables systematic data collection, which powers algorithmic decision-making processes, ultimately manifesting in user-facing services. Our analysis reveals that successful implementations require balanced development across all layers, with particular attention to integration mechanisms between them. The framework contributes to both theory and practice by bridging previously disconnected domains of digital government research, identifying critical dependencies that influence implementation success, and providing a structured approach for analysing the maturity and development pathways of AI-enabled government systems.

# 1. The Rise of the AI-Enabled State

The digital transformation of the public sector has evolved through several distinct phases over the past three decades (Dunleavy and Margetts 2023; Lemke et al. 2020; Millard, 2023; Mountasser & Abdellatif, 2023). From basic *e-government* initiatives focused on service digitisation through to *smart government* solutions leveraging sensor networks and real-time data, today's *AI-enabled digital transformation* builds upon these foundations, with prospect to fundamentally reshape government operations, public service delivery, and policy-making through advanced data science and artificial intelligence capabilities (Engin & Treleaven, 2019; Ubaldi et al., 2019; van Noordt & Misuraca, 2022; Straub et al., 2023; Iosad et al., 2024).

As governments worldwide implement AI technologies, they face significant challenges integrating these innovations into coherent systems. Current implementation approaches often treat technological elements as separate domains with distinct governance frameworks, leading to implementation failures, governance gaps, and missed opportunities to leverage AI's full potential for public value creation. This fragmentation becomes particularly problematic as AI capabilities advance, especially with the rapid development of foundation models (Zhou et al., 2024; Yang et al., 2023) and generative AI (Feuerriegel et al., 2023).

Four techno-political concepts have become central to modern public sector development:

- **Digital Public Infrastructure (DPI)** - shared foundational digital systems established in the public interest to enable scalable innovations and delivery of public and private services to end-users.
- **Data-for-Policy (DfP) -** systematic collection, analysis, and application of data to inform, implement, and evaluate public sector operations and policy decisions, leveraging AI capabilities for enhanced insights.
- **Algorithmic Government / Governance (AG) -** integration of algorithmic systems and AI-enabled processes, including artificial agents, to augment or automate public sector operations and decision-making while maintaining human oversight.
- **GovTech -** digital solutions and applications built to modernise government operations through public-private collaboration, enabling efficient and innovative service delivery.

Existing literature tends to treat these concepts as separate trends or parallel developments. Digital Public Infrastructure (DPI) is often studied through the lens of platform governance and digital commons (O'Reilly, 2010; World Bank, 2024; Eaves et al., 2024), while Data-for-Policy (DfP) research focuses on evidence-based policymaking and data analytics capabilities (Engin et al., 2024; Dabalen et al., 2024; Maffei et al., 2020; Rubinstein et al., 2016). Similarly, Algorithmic Government / Governance (AG) is typically examined through public sector automation and AI governance perspectives (Gritsenko & Wood, 2022; Engin & Treleaven, 2019; Yeung, 2018), while GovTech is primarily discussed in terms of innovation ecosystems and digital service delivery (Bharosa, 2022; Nii-Aponsah et al., 2021; Nose, 2023).

This paper introduces the Algorithmic State Architecture (ASA), a novel framework that conceptualises these elements not as isolated developments but as interdependent layers of an emerging AI-enabled state. By structuring these concepts as a *Foundational Layer* (Digital Public Infrastructure), *Intelligence Layer* (Data-for-Policy), *Process Layer* (Algorithmic Government / Governance), and *Service Layer* (GovTech), we reveal the systematic relationships and enabling mechanisms between them. This layered architecture demonstrates how each level builds upon and enables the capabilities of others, creating a coherent whole that is greater than the sum of its parts.

Our contributions to understanding AI-enabled government transformation are threefold:

1. We demonstrate how these four elements form a coherent system with specific enabling relationships and feedback mechanisms between layers
2. We identify critical dependencies and development pathways that influence the maturity and effectiveness of AI adoption in government
3. We provide a structured approach for analysing how foundational infrastructure enables data collection and analysis, which powers algorithmic processes that manifest in innovative service delivery

Our analysis draws on multiple case studies across different jurisdictional contexts (Estonia, Singapore, India, and the UK) to illustrate how this architecture manifests in practice. These cases reveal both the potential and challenges of implementing such a layered approach, highlighting key considerations for policymakers and practitioners. They demonstrate that successful AI adoption requires strategic coordination across all layers, with particular attention to scalability requirements and security considerations at each level. The cases also reveal how weaknesses in foundational elements often constrain possibilities at higher layers, creating implementation bottlenecks that limit overall effectiveness.

The ASA framework bridges previously disconnected domains of digital government research, identifies critical dependencies that influence implementation success, and offers a maturity model that enables organisations to assess their current state and plan strategic pathways for advancement.

The Algorithmic State Architecture exists within a complex techno-political ecosystem that extends beyond the four core elements we focus on in this paper. As illustrated in Figure 1, this broader ecosystem includes external influences such as international competition and cooperation, AI technology companies, and citizen engagement that continuously shape government AI implementation. Additionally, the core architecture operates within surrounding operational dimensions including knowledge translation between AI advancements and policy practice, business and innovation models spanning proprietary and open approaches, and implementation scales from global to individual. These elements are further contextualised by higher-level enabling, governance, and principles layers that provide the institutional frameworks, legal boundaries, and ethical foundations for AI-enabled government. While acknowledging the importance of these broader dimensions, this paper deliberately focuses on the four core architectural elements—DPI, DfP, AG, and GovTech—to develop a cohesive understanding of their interdependencies and relationships. Future research might explore how these core elements interact with the broader ecosystem components identified in our comprehensive framework.

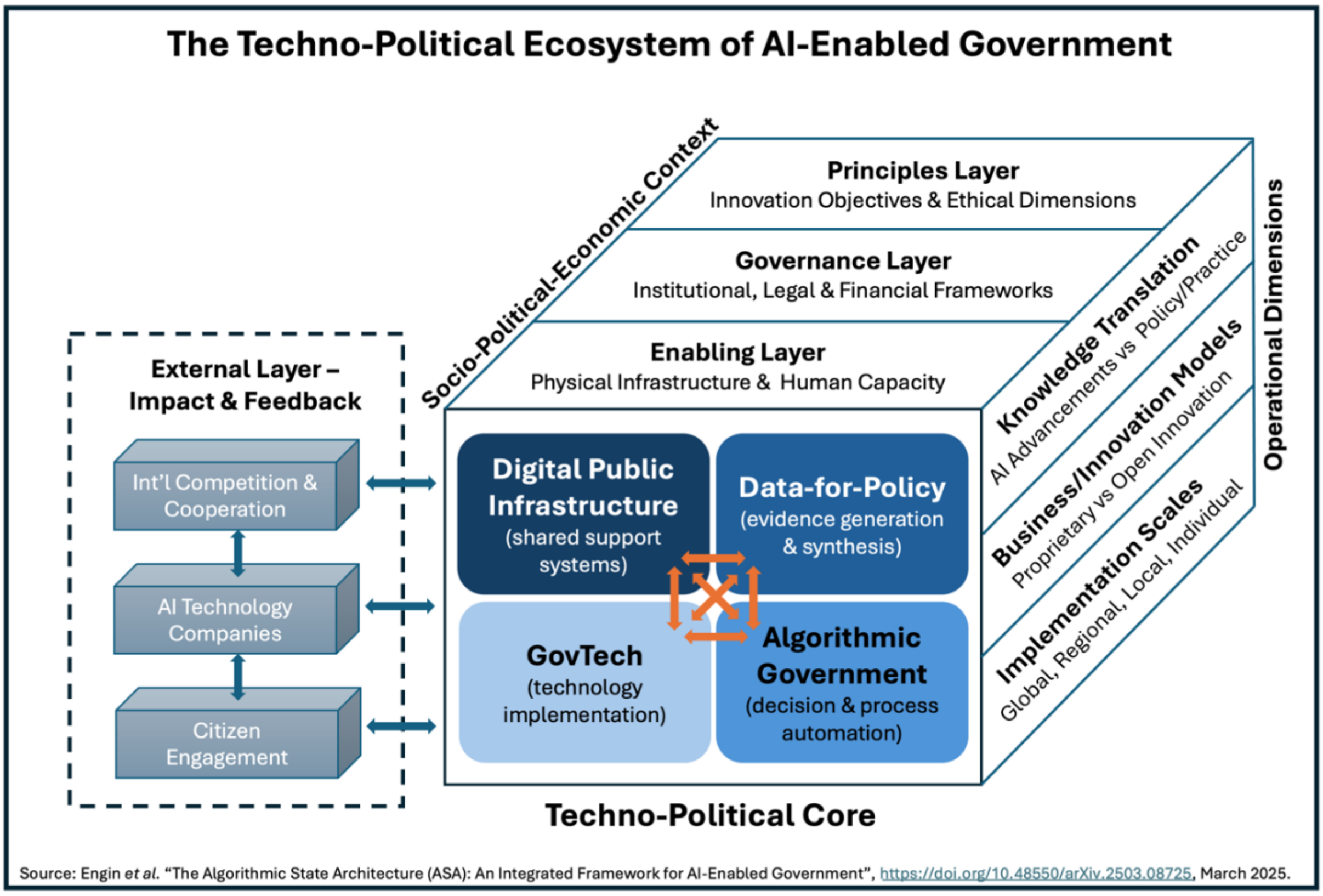

*Figure 1: The Algorithmic State Architecture exists within a complex techno-political ecosystem including the socio-political-economic context, operational dimensions, and external influences that continuously shape government AI implementation. The core ASA elements (DPI, DfP, AG, and GovTech) interact with surrounding governance frameworks, enabling conditions, and broader principles.*

The remainder of this paper is structured as follows. Section 2 reviews the evolution of digital government and situates our four key concepts within existing literature. Section 3 introduces the Algorithmic State Architecture in detail, examining the specific mechanisms and relationships between layers. Section 4 presents our case study analysis, demonstrating how the framework helps understand real-world implementations. Section 5 discusses implications for policy and practice, while Section 6 concludes with future research directions.

# 2. Theoretical Foundations: From Digital Government to AI-Enabled Transformation

Existing frameworks for government digital transformation span several distinct approaches but have not fully addressed the integration challenges of AI-enabled government. Digital Era Governance (DEG) (Dunleavy et al., 2006; Dunleavy & Margetts, 2023) emphasises organisational reintegration but provides limited guidance on AI implementation. Government-as-a-Platform (GaaP) approaches (O'Reilly, 2010; Pope, 2019) offer robust technical architectures but prioritise infrastructure over governance mechanisms. Maturity models (Hujran et al., 2021; Abu Bakar et al., 2020; Gil-García et al., 2016) attempt to map transformation through stages but oversimplify the complex interplay between technical and policy considerations. Meanwhile, international frameworks like the UN E-Government Survey (United Nations, 2020), World Bank's Digital Government Readiness Assessment Toolkit (World Bank, 2020), and OECD Digital Government Framework (OECD, 2020) provide comprehensive coverage but have yet to fully incorporate AI-specific challenges.

The implementation of artificial intelligence in public sector development represents a fundamental shift in how governments operate, make decisions, and deliver services to citizens (Engin & Treleaven, 2019; van Noordt & Misuraca, 2022; Straub et al., 2023; Iosad et al., 2024). AI's capacity to process vast amounts of unstructured data, seemingly understand and generate human language, analyse visual and spatial information, optimise complex systems, and rapidly adapt to new contexts presents unprecedented opportunities for public sector transformation. However, these capabilities also embody inherent techno-political characteristics through their opacity, robustness and reliability challenges, embedded biases, limitations in representing complex social realities, and questions of artificial agency in public decision-making (Samson et al., 2023) .

This section examines how four key techno-political elements—Digital Public Infrastructure, Data-for-Policy, Algorithmic Government/Governance, and GovTech—have evolved as distinct yet interrelated forces in public sector development. We analyse each element's conceptual foundations, technical implementations, and governance implications, revealing both their individual significance and the critical gaps that emerge when they are treated in isolation.

## 2.1. Digital Public Infrastructure: From e-government platforms to digital commons

Digital Public Infrastructure (DPI) comprises *shared foundational digital systems established in the public interest to enable scalable innovations and delivery of public and private services to end-users.* Operating as an intermediate layer between physical infrastructure (broadband networks, computing hardware, data centres, etc.) and sectoral applications (social protection systems, public services, digital payments, etc.), DPI creates a dynamic bridge between physical and digital worlds (World Bank, 2024; Eaves et al., 2024).

The concept has evolved from basic digital connectivity initiatives to sophisticated shared systems, marked by key developments: O'Reilly's Government-as-a-Platform framework (2010)(Kuhn et al., 2023), India Stack's implementation demonstrating viability (Alonso et al., 2023), and formal recognition as distinct infrastructure. Recent developments centre on AI integration and sovereignty considerations (Nagar & Eaves, 2024) (UNDP, 2023).

DPI implementations span multiple technical approaches: microservices-based architectures utilising standardised APIs, cloud computing raising data sovereignty questions (Couture & Toupin, 2019; Kushwaha et al., 2020), distributed data infrastructure addressing interoperability (Tan et al., 2022; Bokolo, 2022), blockchain protocols enabling trustless transactions (Hartelius, 2023),  and digital commons frameworks supporting shared computational resources (Dulong de Rosnay & Stalder, 2020). Digital twins and cyber-physical systems function as critical interface layers within DPI, bridging physical infrastructure with digital capabilities  (Jones et al., 2020; Bennett et al., 2023; Pias et al., 2025).

AI integration fundamentally enhances infrastructure capabilities while introducing new considerations. Machine learning enables advanced identity verification (Minaee et al., 2023), natural language interfaces (Makasi et al., 2022), intelligent API orchestration (Panchal et al., 2024; Charankar & Pandiya, 2024), and automated data exchange protocols (Spanaki et al., 2021). However, these capabilities embed specific limitations—from demographic biases in biometric recognition to linguistic biases in interface design. Foundation models reshape DPI architecture through demands for specialised hardware, distributed training networks, and edge deployment frameworks, introducing critical questions of model and infrastructure sovereignty (Samson et al., 2023).

Implementation factors increasingly centre on architectural pattern selection (monolithic vs. microservices), deployment models (public/private/hybrid cloud), data architecture specifications (centralised/distributed databases), and AI compute distribution approaches. The path forward requires frameworks that balance system optimisation with governance requirements while addressing inherent limitations and biases of AI systems in serving diverse social realities and maintaining robust security postures.

## 2.2. Data-for-Policy (DfP): Evolution of evidence-based policy-making

Data-for-Policy (DfP) encompasses the *systematic collection, analysis, and application of data to inform, implement, and evaluate public sector operations and policy decisions, leveraging AI capabilities for enhanced insights*. The concept (Engin et al., 2024; Dabalen et al., 2024; Maffei et al., 2020; Rubinstein et al., 2016) represents a holistic interpretation of an established field, evolving from social statistics (Bauer, 1966) and evidence-based policy (Davies et al., 2000), through the big data revolution in the public sector (Kitchin, 2014; Mergel et al., 2016; Vichi & Hand, 2019; Engin et al., 2020), to the current integration of AI-enabled capabilities  (Athey, 2017; Pencheva et al., 2018). The DfP concept is particularly relevant in characterising the contemporary landscape where data-driven and data-centric AI systems are increasingly embedded in policy processes (Zha et al., 2025; Jarrahi et al., 2023; Charles et al., 2022), enabling sophisticated analysis and automated decision support while introducing new considerations for governance and oversight.

National and international implementations demonstrate diverse approaches: the US Census Bureau's integration of administrative records and machine learning for population estimation (US Census Bureau, 2024), Cochrane's AI-enabled systematic review tools (Metzendorf & Klerings, 2025), the World Bank's Data Catalog (World Bank), Estonia's X-Road data exchange platform (X-Road), and China's City Brain systems for urban management (Xu et al., 2024). These implementations illustrate a range of technical architectures and governance models that can support data-driven policy development, each tailored to specific institutional contexts and policy objectives.

DfP encompasses comprehensive technical infrastructure including official statistical systems (Allin, 2019; Radermacher, 2019; Groshen, 2021), data pipeline architectures (batch/streaming) (Sresth et al., 2023), analytical frameworks (descriptive/predictive/prescriptive) (Segui et al., 2025), data integration patterns (ETL/ELT) (Walha et al., 2024), and emerging computational paradigms (distributed analytics, federated learning, privacy-preserving computation) (Chen et al., 2024) (Zhang et al., 2021). Contemporary data architectures have evolved from centralised data warehouses to distributed data mesh architectures (Machado et al., 2021) and federated analytics networks (Wang et al., 2022), each addressing specific needs in data governance and computational distribution.

AI integration transforms these architectures across the data lifecycle (Kumar et al., 2024; Sable et al., 2023). In data collection, AI enables intelligent capture from unstructured sources and automated quality assessment. For analytics, AI extends traditional methods through deep learning for pattern recognition, NLP for textual analysis, and computer vision for spatial data. In evidence synthesis, language models facilitate literature screening, data extraction, and risk assessment (Qureshi et al., 2023; van de Schoot et al., 2021). Foundation models introduce new analytical possibilities from zero-shot learning to few-shot adaptation (Meshkin et al., 2024). However, these capabilities introduce critical considerations around bias and representation - from demographic skews in training data to cultural assumptions in language models (Crawford, 2021).

Critical implementation factors include data architecture selection (warehouse/lake/mesh), analytical deployment models (batch/real-time/hybrid), computation distribution approaches (centralised/distributed/federated), and AI model governance frameworks. Implementation choices between human judgment and algorithmic analysis reflect deeper questions about expertise and authority in policy-making (Guerrero & Margetts, 2024). The path forward requires frameworks that balance analytical power with governance requirements while addressing AI systems' inherent limitations in capturing complex social realities for policy decisions.

## 2.3. Algorithmic Government / Governance (AG): AI agency in decision-making

Algorithmic Government / Governance (AG) refers to *the integration of algorithmic systems and AI-enabled processes, including artificial agents, to augment or automate public sector operations and decision-making while maintaining human oversight.* The evolution of the concept (Engin & Treleaven, 2019; Yeung, 2018; O'Reilly, 2013; Gritsenko & Wood, 2022; Aneesh, 2009) spans from pre-digital bureaucratic rationalisation to today's AI-driven decision systems(Newman et al., 2022), raising fundamental questions about how democratic state functions will transform as artificial agents increasingly impact high-stakes decision making (Sidhu et al., 2024; O'Callaghan, 2023; Grimmelikhuijsen & Meijer, 2022; Levy et al., 2021)(Crawford, 2021).

A foundational discussion in AG research is to establish the roles and interactions of human and algorithmic agents in decision-making processes (Chong et al., 2021; Engstrom et al., 2020; Longoni et al., 2023; Erkkilä, 2024; Alon-Barkat & Busuioc, 2024; Engin, 2025). This agency exists along a spectrum that reflects different relationships between human governance and algorithmic systems as follows (Engin, 2022; Parycek et al., 2024; Campbell-Verduyn et al., 2016):

**Government / Governance *through* algorithms:** Humans maintain primary agency while algorithms serve as tools executing defined tasks under direct supervision (Yang & Zhu, 2024; Ompusunggu et al., 2021), such as automated tax processing systems that apply predefined rules to standardised inputs. The cognitive division of labor heavily favors human expertise, with algorithms handling bounded, repetitive processes. Even as AI capabilities advance, this approach deliberately constrains algorithmic authority through tight procedural boundaries and human oversight requirements.

**Government / Governance *by* algorithms**: This shifts significant decision-making agency to algorithmic systems (Acharya et al., 2025), as seen in predictive policing platforms that autonomously determine patrol allocations (Wilson, 2019) or autonomous AI agents operating independently within complex urban digital infrastructure (Hintze & Dunn, 2022). These systems transform from passive tools to active partners through their capacity to process vast datasets and identify patterns beyond human capabilities. Humans often assume supervisory roles (Green, 2022), reviewing algorithmic recommendations rather than generating solutions. This redistribution of cognitive labour challenges conventional accountability structures as decision influence becomes more distributed across human-machine systems, particularly when algorithmic agents operate continuously and make interconnected decisions within complex cyber-physical infrastructures.

**Government / Governance *with* algorithms:** This balanced partnership aims to leverage complementary strengths of both human and algorithmic agents (Collins et al., 2024; Pflanzer et al., 2023), exemplified by urban planning platforms that model policy impacts while human planners incorporate insights into deliberative processes. Various theoretical approaches structure this collaboration (Bullock et al., 2022; Geng & Varshney, 2022; Engin & Hand, 2025.a), ranging from models that alternate decision authority based on context, to frameworks that establish systematic validation checkpoints, to approaches that integrate algorithmic analysis into multi-stakeholder deliberations. Human-algorithm partnerships in governance require flexibility in how decision authority and processes are distributed across the socio-technical system.

**Government / Governance *of* algorithms:** This meta-level approach examines how the human-machine relationship itself is governed (Díaz-Rodríguez et al., 2023, Esposito & Tse, 2024), as demonstrated by Canada's Directive on Automated Decision-Making (Government of Canada, 2021), which requires impact assessments based on potential effects on rights, or the EU's Artificial Intelligence Act (European Union, 2024) with its risk-based regulatory framework. It focuses on aligning algorithmic operations with human missions and values, ensuring appropriate transparency, and creating mechanisms for contesting algorithmic decisions.

From a technical perspective, AG focuses on computational architectures enabling AI agency in governance decisions while managing the distribution of decision authority through interaction protocols and oversight mechanisms (Khowaja et al., 2025; Xiong et al., 2022). These systems distribute decision authority between human and machine agents through reasoning engines, machine learning models, and hybrid frameworks that implement varying levels of autonomy (Ren et al., 2023; Kierner et al., 2023; Morris et al., 2024; Zhang & Meng, 2024).

AG systems incorporate AI capabilities calibrated for public administration—from NLP and computer vision to reinforcement learning - which necessitates specialised safeguards including transparency and explainability mechanisms (Arrieta et al., 2020; Ribeiro et al., 2016; Rudin, 2019), fairness frameworks (Barocas et al., 2023; Kleinberg et al., 2016), accountability mechanisms (Raji et al., 2020; Koshiyama et al., 2022), and technical robustness measures including adversarial testing protocols (Hannon et al., 2023; Kurakin et al., 2018), distribution shift detection (Rabanser et al., 2019), and formal validation and verification techniques (Seshia et al., 2022; Wotawa, 2021) to ensure reliable operation.

As these systems evolve, the key concern is to incorporate democratic oversight and safeguards that maintain legitimacy while leveraging AI capabilities (Engin & Hand, 2025.b; Crawford, 2021; Pasquale, 2016; Kitchin, 2016), addressing the unique requirements of algorithmic agency in government while preserving accountability.

## 2.4. GovTech: The rise of public-private digital innovation

GovTech encompasses *digital solutions and applications built to modernise government operations through public-private collaboration, enabling efficient and innovative service delivery*. The contemporary understanding of this concept (European Union, 2022; Bharosa, 2022; OECD, 2024; Nose, 2023; Diakite & Wandaogo, 2024) emerged around 2015-2016 with the establishment of GovTech Singapore, expanding through specialised venture funds like PUBLIC (PUBLIC, 2017) and The CivTech Alliance. The COVID-19 pandemic accelerated GovTech adoption (Dener et al., 2021), while recent evolution has seen maturation into an integrated approach encompassing procurement reform, standardised frameworks, and marketplace development (World Economic Forum & Berlin Government Technology Centre, 2025).

While GovTech shares boundaries with related domains like RegTech (Bolton & Mintrom, 2023), LegalTech (Mania, 2023), HealthTech (Vincent, 2025), EdTech (Kerssens & van Dijk, 2022) , and CivicTech (Saldivar et al., 2019), it maintains distinctiveness through its explicit focus on government operations modernisation and structured public-private collaboration frameworks. Additionally, these domains increasingly intersect with GovTech in practice - for instance, when RegTech solutions are adopted by government regulatory bodies, when LegalTech platforms are integrated into judicial systems, or when CivicTech applications interface with government service delivery. This convergence reflects the broader trend toward integrated digital ecosystems where technological solutions often serve multiple purposes across public and private sectors.

GovTech encompasses technical platforms and solution architectures that enable innovation in government operations. Notable examples include Singapore's LifeSG app for integrated citizen services, the UK's GOV.UK Notify platform, and Australia's BuyICT platform for streamlined procurement. The ecosystem includes development platforms (low-code/no-code), innovation infrastructure (sandboxes/prototyping), market enablers (procurement/vendor management), integration frameworks (APIs/connectors), and ecosystem tools (partnership platforms/innovation hubs).

Implementation architectures manifest through multiple technical approaches. Denmark's Borger.dk demonstrates a successful digital service platform utilising modern application architectures. New Zealand's SmartStart exemplifies cloud-native architecture enabling life-event services, while Singapore's Parking.sg showcases robust mobile-first implementation. These implementations can be deployed through government cloud platforms, hybrid infrastructure models, or distributed service architectures.

AI integration is enhancing these capabilities while introducing new techno-political dynamics. Machine learning operations enable automated service optimisation, intelligent API management supports dynamic service composition, and automated testing frameworks ensure service reliability. The emergence of foundation models introduces new service possibilities but also concerns about dependency on private AI infrastructure and the balance of public-private control over critical government services.

Critical implementation factors include platform architecture selection (monolithic/microservices/serverless), deployment models (cloud/hybrid/edge), integration patterns (synchronous/asynchronous/event-driven), and development framework specifications. Technical trajectories indicate evolution toward AI-enabled service platforms, but these developments raise fundamental questions about public sector autonomy and market power in digital government. The path forward requires frameworks that balance innovation with governance requirements while addressing the implications of increasing private sector AI capability for public sector independence and democratic accountability.

## 2.5. Gap Analysis: Need for Integrated Framework

While existing research and grey literature provides extensive insight into individual elements of AI-enabled government transformation, current approaches reveal several critical gaps that the ASA framework addresses.

First, the fragmentation of research across DPI, DfP, AG, and GovTech creates artificial boundaries that obscure their fundamental interdependencies. This separation produces significant implementation problems: discussions of algorithmic governance often overlook how DPI capabilities enable or constrain AI implementation; governance frameworks fail to address end-to-end accountability across layers; misaligned investment strategies lead to advanced capabilities in one domain without sufficient foundation in others; and limitations in one area (such as data quality) fundamentally constrain possibilities in others regardless of technological sophistication.

Second, existing frameworks tend to either overemphasise technical aspects while understating governance implications (as in GaaP approaches), or focus on governance without fully addressing technical implementation requirements (as seen in algorithmic accountability frameworks). This dichotomy fails to capture the techno-political nature of AI-enabled government transformation, where technical choices inherently embed governance implications and vice versa, leading to solutions that appear technically sound but face governance challenges in practice, or governance frameworks that prove impractical to implement.

Third, current approaches largely treat AI integration as an enhancement to existing systems rather than recognising how it fundamentally reshapes the relationships between infrastructure, data, decision-making, and service delivery. The emergence of foundation models, in particular, creates new dependencies and capabilities that transcend traditional boundaries between these domains, requiring governments to reconsider integration patterns, data flows, and decision authority distributions across previously separate systems.

Fourth, existing maturity models and implementation frameworks typically present linear progression paths that inadequately capture the complex interactions and feedback loops between different elements of AI-enabled government. This oversimplification can lead to implementation strategies that fail to account for critical dependencies and potential bottlenecks, resulting in stalled projects, stranded capabilities, and unrealised public value.

These gaps point to the need for an integrated framework that: **1)** recognises the systematic relationships between different elements of AI-enabled government, **2)** addresses both technical and governance considerations as inherently linked concerns, **3)** accounts for how AI capabilities reshape traditional boundaries and relationships, and **4)** captures the complex dynamics and dependencies that influence implementation success.

The Algorithmic State Architecture (ASA) framework introduced in the next section addresses these gaps by conceptualising these elements as interdependent layers of a coherent system, providing a structured approach to understanding and implementing AI-enabled government transformation.

# 3. The Algorithmic State Architecture (ASA): An Integrated Framework

The Algorithmic State Architecture (ASA) represents a novel approach to conceptualising and implementing AI-enabled government transformation. While existing frameworks tend to treat digital public infrastructure, data-driven capabilities, algorithmic processes, and service delivery as parallel developments, ASA positions them as interdependent layers of a coherent system. This section details the framework's structure, mechanisms, and dynamics.

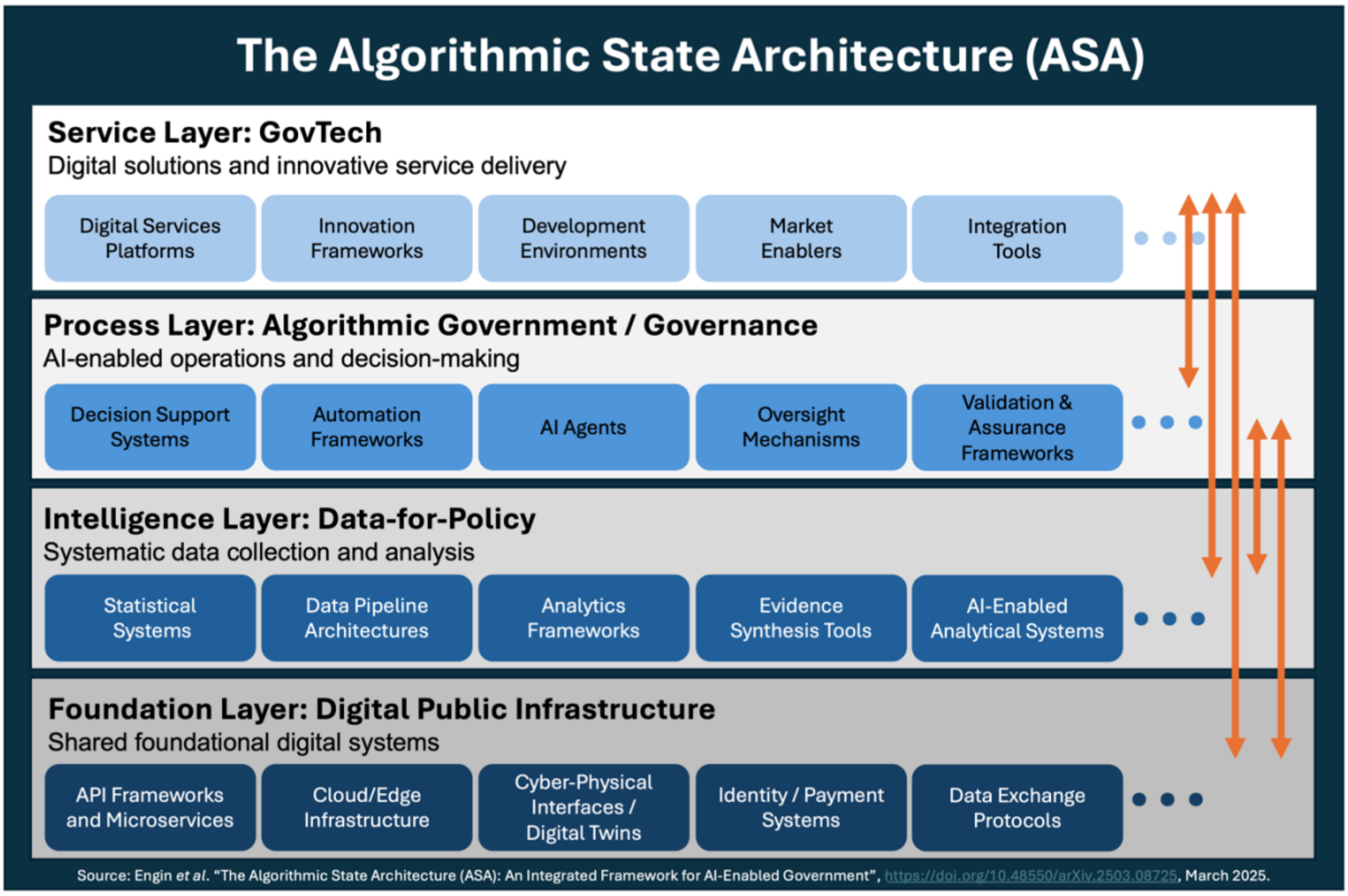

*Figure 2: The ASA framework conceptualises AI-enabled government transformation through four interconnected and scalable layers: Foundation Layer (Digital Public Infrastructure), Intelligence Layer (Data-for-Policy), Process Layer (Algorithmic Government), and Service Layer (GovTech). Bidirectional arrows indicate the enabling relationships and feedback mechanisms between different layers, where capabilities at one layer both enable and constrain possibilities at others.*

## 3.1. Framework Overview and Design Principles

The ASA framework conceptualises AI-enabled government transformation through four interconnected and scalable layers that bridge technical capabilities with governance requirements as seen in Figure 2. While these four elements - DPI, DfP, AG, and GovTech - exist as interconnected components within a broader techno-political ecosystem (Figure 1), our analysis reveals that they also form a structured architecture with specific enabling relationships and dependencies. Figure 2 presents the Algorithmic State Architecture as a layered framework that illustrates how foundational digital infrastructure enables systematic data collection and analysis, which powers algorithmic decision-making processes, ultimately manifesting in user-facing government services.

This approach draws inspiration from layered architectures in computing systems, notably the OSI model (Day, 2008), while acknowledging the inherent tensions in strictly separated layers (Dourish, 1996). As observed in the evolution of computing architectures, we recognise that while conceptual separation provides analytical clarity and governance boundaries, practical implementations often develop "cross-layer optimisations" or even "layer violations" for efficiency and innovation. The ASA framework anticipates that AI-enabled government implementations will similarly balance the benefits of clear layer separation with the pragmatic requirements for cross-layer interaction, particularly as AI capabilities evolve to create new integration opportunities. This tension between modular isolation and cross-layer flexibility informs our design principles, which aim to maintain architectural integrity while allowing strategic adaptability in implementation.

The ASA framework is built on six key design principles that govern both its conceptual structure and practical implementation:

1. *Layer independence and sovereignty* represents how each layer maintains operational autonomy and governance sovereignty. This enables independent evolution and improvement while adhering to standardised interfaces, preserves democratic control and accountability at each level, and allows for context-specific implementation choices.
2. *Systemic interdependence* acknowledges that layers exhibit strong functional dependencies where capabilities enable and constrain possibilities. Technical choices in one layer shape governance options in others, infrastructure capabilities influence service possibilities, and data capabilities enable and constrain algorithmic choices.
3. *Bidirectional feedback* recognises that information and influence flow both upward (enabling) and downward (informing) through the architecture. This creates dynamic feedback loops between technical implementation and governance requirements, enables continuous adaptation and learning across layers, and supports iterative development and improvement.
4. *Progressive maturity* allows layers to develop at different rates while maintaining minimum viable capabilities. This principle supports context-specific development priorities, enables incremental capability building, and allows for evolutionary implementation paths tailored to local needs and resources.
5. *Techno-political integration* emphasises that technical choices and governance requirements are inherently linked. Infrastructure decisions embed political choices, algorithmic systems reflect governance values, and service design shapes citizen-state relationships. This principle recognises the inseparable nature of technical implementation and governance considerations.
6. *Democratic accountability* ensures the preservation of human oversight and democratic control throughout the system. This principle maintains transparency across layers, enables public scrutiny and intervention, and ensures alignment with public values and democratic principles.

These principles collectively recognise that AI integration fundamentally reshapes relationships between layers, technical and governance considerations are inseparable, implementation choices have broader societal implications, and democratic oversight must be maintained throughout the system. The principles guide both the conceptual understanding of how these layers interact and the practical implementation of AI-enabled government transformation.

## 3.2. Layer-specific Mechanisms

The ASA framework comprises four distinct layers, each characterised by specific mechanisms that enable AI-enabled government transformation. This section details the core functions, key components, success factors, and constraints for each layer.

### 3.2.1. Foundation Layer: Digital Public Infrastructure

The *Foundation Layer* provides shared foundational digital systems that bridge physical and digital infrastructure while enabling scalable innovations and service delivery. Its core functions include facilitating bi-directional information flows and establishing base technological capabilities for higher layers. Key components of this layer include API frameworks and microservices architectures, cloud and edge infrastructure, digital twin systems, identity and payment systems, data exchange protocols, and cyber-physical interfaces. These components create the technical foundation upon which other layers build their capabilities. Success factors for the *Foundation Layer* include interoperability and standardisation across systems, scalability to handle growing demands, security to protect critical infrastructure, sovereignty over key digital assets, effective physical-digital integration and resilience to change. This last factor—the ability to accommodate technological advances without disrupting essential services or requiring complete system redesign—is exemplified by Estonia's "no legacy" approach, which enables continuous modernisation while maintaining operational continuity. These factors determine the layer's ability to support higher-level functions effectively. The layer faces several constraints, including the management of technical debt from legacy systems, infrastructure costs associated with maintenance and upgrades, challenges in maintaining sovereignty over critical infrastructure components, and ongoing security vulnerabilities that must be addressed.

### 3.2.2. Intelligence Layer: Data-for-Policy

The *Intelligence Layer* enables systematic data collection, integration, and analysis to support evidence-based governance. Its core functions encompass comprehensive data management, analysis and insight generation, evidence synthesis and evaluation, and policy impact assessment. Key components include statistical systems for data collection and processing, data pipeline architectures for efficient data flow, analytics frameworks spanning descriptive to prescriptive capabilities, evidence synthesis tools, and AI-enabled analytical systems. These components work together to transform raw data into actionable policy insights. Success factors include maintaining high data quality and coverage, developing sophisticated analytical capabilities, ensuring insights are actionable for policy-making, maintaining methodological rigour, and preserving privacy throughout data operations. The layer confronts constraints related to data availability and quality, strict privacy requirements, analytical complexity in processing diverse data types, potential bias in both data and models, and challenges in maintaining interpretability of advanced analytical systems.

### 3.2.3. Process Layer: Algorithmic Government

The *Process Layer* implements AI-enabled decision-making and algorithmic automation while maintaining appropriate human oversight. Its core functions include process automation, human-algorithm collaboration, and governance of algorithmic systems. Key components encompass decision support systems, automation frameworks, AI agents, oversight mechanisms, explainability tools, and validation frameworks. These components enable the systematic integration of AI capabilities into government operations while maintaining accountability. Success factors include achieving process optimisation while maintaining quality, establishing effective human oversight, ensuring algorithmic transparency, and maintaining democratic accountability throughout automated processes. Critical to these success factors are well-designed transfer mechanisms that enable easy handover to human operators when systems encounter edge cases or unusual circumstances. The layer must address constraints including algorithm bias, transparency requirements, accountability mechanisms, appropriate distribution of agency between human and machine actors, and maintenance of democratic control over automated processes.

### 3.2.4 Service Layer: GovTech

The *Service Layer* focuses on delivering user-facing services and enabling innovation through public-private collaboration. Its core functions include digital service delivery, ecosystem development, and service modernisation. Key components include digital service platforms, innovation frameworks, development environments, market enablers, and integration tools. These components support the delivery of government services while fostering innovation. Success factors encompass user adoption of digital services, maintaining high service quality, developing a vibrant innovation ecosystem, enabling effective public-private collaboration, and achieving seamless service integration. Drawing from frameworks like the Office for Statistics Regulation's approach, additional critical success factors include trustworthiness (the confidence in those producing models that are to be used in the public domain), quality (data and methods that produce assured outputs), and value (demonstrable public benefit) (OSR, 2022). The layer faces constraints related to user acceptance of digital services, addressing the digital divide, building innovation capacity, managing private sector dependencies, and balancing market power dynamics in service delivery.

## 3.3. Inter-layer Relationships and Cross-Cutting Requirements

While each layer of the ASA framework maintains operational independence, the system's effectiveness depends on relationships between layers, manifesting through vertical dependencies, cross-layer effects, and system dynamics. Understanding these relationships reveals bottlenecks, highlights strategic investment opportunities, and shows how technical choices and governance requirements propagate throughout the system.

The primary relationships occur between adjacent layers, creating a chain of dependencies:

- ***Foundation → Intelligence***: Infrastructure enables data collection and processing
- ***Intelligence → Process***: Data insights power algorithmic decision-making
- ***Process → Service***: Automated processes enable service delivery
- ***Service → Foundation***: Service requirements shape infrastructure needs

These vertical dependencies form the framework's backbone, establishing how capabilities and constraints flow through the system. Beyond these direct relationships, significant interactions also occur between non-adjacent layers through indirect pathways:

- ***Foundation ↔ Process***: Infrastructure capabilities influence algorithmic possibilities
- ***Intelligence ↔ Service***: User interactions generate data; insights shape services
- ***Foundation ↔ Service***: Infrastructure enables and constrains service innovation
- ***Intelligence ↔ Process***: Algorithmic processes generate and consume data

These cross-layer effects demonstrate how the ASA framework functions as an integrated system rather than a simple hierarchy.

The ASA framework directly addresses the persistent challenge of organisational silos in digital transformation. Traditional government structures often separate technical implementation from policy development, creating situations where managers may not understand technical limitations and developers may not appreciate governance requirements. This silo mentality has contributed to implementation failures across sectors—from the 2008 financial crisis to numerous government IT projects. By explicitly mapping dependencies between layers, the ASA framework creates a shared conceptual model that bridges technical and policy domains. It establishes clear interfaces that enable effective communication while preserving necessary specialisation, helping organisations develop governance mechanisms that align technical possibilities with policy objectives. This integrated perspective is particularly important for AI implementation, where the complexity of systems makes siloed approaches especially problematic.

The interaction between these dependencies and effects creates broader system dynamics that influence how the framework evolves: *feedback loops* where service usage informs process optimisation; *adaptation mechanisms* where infrastructure evolves based on service needs; *innovation pathways* where cross-layer interactions enable new capabilities; and *constraint propagation* where limitations in one layer affect others.

Scalability and security span the entire architecture as cross-cutting requirements. Scalability manifests differently across layers: the *Foundation Layer* requires distributed computing models and elastic infrastructure; the *Intelligence Layer* needs distributed processing architectures; the *Process Layer* must maintain accuracy and oversight as decision volumes increase; and the *Service Layer* emphasises performance as user adoption grows. Cross-layer scalability dynamics reveal critical interdependencies where bottlenecks in one layer can constrain the entire system.

Security requires complementary measures across all layers: the *Foundation Layer* establishes the security baseline through infrastructure protection; the *Intelligence Layer* focuses on data protection and privacy-preserving methods; the *Process Layer* addresses algorithmic integrity; and the *Service Layer* emphasises user protection through secure interfaces. These requirements create both tensions (authentication bottlenecks, slower innovation cycles) and reinforcements (well-designed API management enhances both security and scalability).

Effective implementation recognises that scalability and security must be addressed holistically across all layers of the ASA framework, with systematic approaches tailored to each development stage—from initial implementation prioritising foundational security, through growth stages requiring careful monitoring, to mature operations focusing on continuous adaptation.

## 3.4. Maturity Model and Development Pathways

The implementation of AI-enabled government transformation through the ASA framework requires understanding how different layers mature and evolve over time. This section presents a streamlined maturity model to help organisations assess capabilities, identify priorities, and plan strategic development.

Each ASA layer has distinct indicators that signal its level of development. The *Foundation Layer* is measured through infrastructure robustness (system reliability and resilience), API coverage (standardisation of service interfaces), and security posture (cybersecurity measures and sovereignty controls). The *Intelligence Layer* is assessed through data quality (completeness, accuracy, timeliness), analytical sophistication (complexity of methods from basic statistics to advanced AI), and insight utilisation (application of analysis in policy and operations). The *Process Layer* is evaluated by automation level (extent of algorithmic processes), decision quality (accuracy and reliability), and oversight effectiveness (robustness of human supervision mechanisms). The *Service Layer* is determined by service coverage (breadth of digital services), user satisfaction (adoption rates and experience quality), and innovation activity (vitality of the GovTech ecosystem).

The ASA framework supports multiple implementation approaches suited to different contexts. *Bottom-up Development* builds strong foundational infrastructure before advancing to higher layers, creating sustainable but potentially slower transformation, as exemplified by Estonia's digital transformation. *Top-down Innovation* starts with service-level initiatives that drive supporting capabilities in lower layers, accelerating visible improvements but requiring careful management of technical debt, as seen in the UK's Government Digital Service. *Hybrid Evolution* involves parallel development across multiple layers with strong coordination mechanisms, balancing benefits of simultaneous progress with increased complexity, illustrated by Singapore's Smart Nation initiative. *Opportunistic Growth* allows context-specific capability development based on immediate needs, recognising that different parts of government may develop at different rates while requiring strong architectural guidance to maintain coherence.

Successful implementation requires attention to several critical factors: resource requirements (technical, human, financial), risk management (security, privacy, operational), change management (stakeholder engagement, capability building), and governance frameworks (oversight, accountability, coordination). The ASA maturity model acknowledges that perfect balance across all layers is rarely achievable. Instead, it emphasises maintaining minimum viable capabilities throughout the architecture while allowing strategic advancement in priority areas based on specific organisational contexts and objectives.

# 4. ASA in Practice: Case Study Analysis

This section applies the Algorithmic State Architecture (ASA) framework to analyse real-world implementations across four jurisdictions. Through comparative analysis, we identify patterns of success, common challenges, and implementation lessons that inform effective AI-enabled government transformation.

Our case study analysis employs a structured comparative approach examining implementations across different contexts, development stages, and governance models. Cases were selected based on three criteria: **(1)** comprehensive implementation spanning multiple ASA layers, **(2)** availability of detailed implementation data, and **(3)** diversity of geographical and governance contexts. The analysis framework assessed each case across all four ASA layers, examining technical implementation approaches, governance mechanisms, and outcomes.

## 4.1. Estonia's X-Road Ecosystem: A Comprehensive ASA Implementation

Estonia's X-Road exemplifies a mature ASA implementation with strong integration across all layers (X-Road; Vassil, 2016; Anthes, 2015; Heller, 2017). At the *Foundation Layer*, the distributed data exchange layer connects decentralised databases while maintaining data sovereignty through the "once-only" principle (Mamrot & Rzyszczak, 2021). Estonia strengthened this foundation with its "Data Embassy" concept—data centres in allied countries that ensure service continuity during disruptions, creating infrastructure-level resilience (Hardy, 2023).

The Intelligence Layer leverages this infrastructure for comprehensive analytics that directly inform policy and operations. The Statistics Estonia integrates information across agencies to produce real-time indicators (e-Estonia, 2018). The Estonian Unemployment Insurance Fund's OTT system analyses over 100,000 client records to predict employment pathways and optimise service delivery, helping civil servants better understand client needs and Prioritise interventions (e-Estonia, 2021). These capabilities are sustained by Estonia's "no-legacy" approach, ensuring continuous modernisation that prevents analytical capabilities from being constrained by outdated systems (Ross, 2022).

The *Process Layer* implements sophisticated algorithmic governance across multiple domains. The automated tax filing system pre-completes tax returns (European

Commission). The e-Court case management system categorises and routes legal cases while maintaining judicial oversight (Adeleye et al., 2022). The cross-border automated tax system with Finland demonstrates algorithmic governance across jurisdictional boundaries, determining tax obligations based on residence time and employment data (EMTA, 2021). Human oversight mechanisms are integrated to ensure accountability without sacrificing efficiency.

At the *Service Layer*, Estonia claims 100% of the government services being delivered digitally as of December 2024 (Kriisa, 2025). Two innovations stand out: the "e-Residency" programme (Hardy, 2023), which extends digital identity and services to non-citizens globally, and the "Personal State" concept (Gołębiowska & Kuczyńska-Zonik, 2024), which customises service delivery based on individual life circumstances. Estonia's Digital Testbed Framework (e-Estonia, 2021) enables controlled experimentation with emerging technologies before full deployment.

Estonia's approach demonstrates both systemic interdependence, where each layer builds on lower capabilities, and progressive maturity as different components evolved at variable rates while maintaining coherence. The robust *Foundation Layer* enabled experimentation at the *Service Layer*, while *Service Layer* innovations drove enhancements to the *Process* and *Foundation Layers*, creating a virtuous cycle of continuous improvement.

## 4.2. Singapore's Smart Nation: Strategic Integration Across ASA Layers

Singapore's Smart Nation initiative demonstrates a comprehensive ASA implementation with strong centralised governance (Sipahi & Saayi, 2024; Singapore Government Developer Portal). The *Foundation Layer* integrates the National Digital Identity system, Singapore Government Technology Stack - SGTS (offering standardised microservices), and Smart Nation Sensor Platform (connecting IoT devices with urban infrastructure). This foundation is strengthened by the Cyber Security Agency's "zero-trust" architecture implementing continuous verification.

The *Intelligence Layer* features GovTech's Data Science and AI Division's cross-agency analytics platforms. Key implementations include the Analytics.gov that supports data analysis and machine learning (ML) initiatives for public agencies, Transcribe - a speech-to-text (STT) - platform that streamlines workflows for public officers, and Urban Planning Analytics platform combining geospatial and demographic data (Smart Nation Singapore). Singapore's TraceTogether contact tracing system exemplified these capabilities during the pandemic (Chow et al., 2023).

The *Process Layer* deploys algorithmic decision-making across multiple domains. The IRAS tax system automatically assesses risk and routes cases accordingly (IRAS), while Singapore's traffic management algorithms adjust signal timing based on real-time conditions (Smart Nation Singapore). The Model AI Governance Framework sets Singapore’s vision for algorithmic accountability and human oversight (Allen et al., 2025).

The *Service Layer* delivers integrated citizen-centric services through applications like LifeSG (providing personalised services based on life events) and the Business Grants Portal (offering unified access to government programmes) (BGP, 2024). The Smart Nation Fellowship Programme fosters public-private collaboration in service innovation (Smart Nation Singapore).

Singapore's approach demonstrates how centralised governance can effectively implement digital transformation when coupled with strategic capability development. Their model emphasises developing technical expertise within government rather than relying solely on the private sector, with GovTech serving as both service provider and capability builder. This sequential development with strong coordination allowed foundational infrastructure to enable increasingly sophisticated capabilities that support innovative service delivery.

## 4.3. India's Aadhaar and India Stack: Strong Foundation with Uneven Layer Development

India's implementation reveals a partial ASA deployment with strengths in specific layers. The *Foundation Layer* centres on Aadhaar, the world's largest biometric identity system, supporting India Stack's API framework with four components (India Stack; Raghavan et al., 2019; Pandey & Chaudhary, 2023; Alonso et al., 2023): Presenceless (identity verification), Paperless (document exchange), Cashless (UPI payment infrastructure), and Consent (data sharing). This infrastructure enables massive scale with relatively low implementation costs.

The Service Layer shows significant innovation through applications like DigiLocker (secure document storage) (Kamath, 2021), UPI (processing over 10 billion monthly transactions) (Government of India), and ONDC (open e-commerce infrastructure) (ONDC). The ecosystem of solutions built on this foundation demonstrates how robust infrastructure can enable service innovation at unprecedented scale.

Meanwhile, the *Intelligence Layer* remains underdeveloped. Despite sectoral initiatives like the National Health Stack (Sharma, 2018), cross-sector data integration and analytics capabilities are limited by inconsistent governance frameworks for data collection and sharing. This constrains potential AI applications despite rich transactional data generated by *Foundation Layer* systems.

The *Process Layer* faces substantial challenges with algorithmic systems for benefit distribution and service eligibility implemented without corresponding frameworks for transparency and oversight (Parsheera, 2024; Larasati et al., 2023; Dattani, 2023). High-profile system failures affecting vulnerable populations highlight the risks of expanding algorithmic processes without adequate safeguards (Amrute et al., 2020).

India's implementation demonstrates ASA's principle of progressive maturity with uneven layer development. The case illustrates how strong *Foundation* and *Service Layers* can deliver significant value even with limited development of intermediate layers, while also highlighting the risks when algorithmic processes expand faster than governance capabilities. India's approach reflects contextual priorities—universal digital access and financial inclusion over sophisticated analytical capabilities—a sequencing that makes sense given its development context but creates challenges as the system evolves towards more advanced AI applications.

## 4.4. United Kingdom's Government Digital Service: Strong Service Innovation with Evolving Infrastructure

The UK's digital government transformation showcases a distinctive ASA implementation with recent developments addressing historical challenges (DSIT & GDS, 2025). The *Service Layer* demonstrates strong user-centred capabilities through GOV.UK's unified platform and the Design System. Key components like GOV.UK Notify and GOV.UK Pay provide reusable infrastructure across departments. The UK's GovTech ecosystem benefits from innovative procurement frameworks including G-Cloud and Digital Marketplace, which have transformed government technology acquisition.

The *Process Layer* shows significant development through robust governance frameworks. The Service Standard ensures user-centred design through formal assessments at key stages. The *Algorithmic Transparency Recording Standard Hub* (2023) has documented systems like the DVSA's MOT Risk Rating algorithm and the Home Office's Complexity Application Routing Solution for visa applications. The Incubator for Artificial Intelligence (i.AI) website showcases various AI-powered solutions designed to support UK Civil Servants. These include products such as Parlex and Lex, which equip policy makers and researchers with AI-driven tools to explore legislation and analyse parliamentary developments, thereby enhancing their capabilities in the policy making process. Another notable tool, Consult, automates the processing of public consultations, further streamlining government operations.

The *Intelligence Layer* presents mixed capabilities. The ONS Data Science Campus and HMRC's Connect system utilise advanced analytics, but the 2025 State of Digital Government Review acknowledges that the current “UK data landscape is not well co-ordinated, interoperable, or enables a unified source of truth” (DSIT & GDS, 2025).

The *Foundation Layer* also remains fragmented and “dependent on decades-old and costly legacy systems with crumbling foundations - large parts of the public sector are still digitally immature” states in the Government 2025 policy paper, *A blueprint for modern digital government* (DSIT & GDS, 2025). At the time of writing, the Government API Catalogue documents 241 APIs representing 34 Departments, though integration remains challenging.

The UK illustrates a "middle-out" development pattern where *Service* and *Process Layers* advanced ahead of comprehensive *Foundation* infrastructure. In January 2025, the UK addressed this fragmentation by merging the Central Digital and Data Office(CDDO), the Geospatial Commission, the original Government Digital Service, the newly established Incubator for Artificial Intelligence (i.AI), and parts of the parts of the Responsible Tech Adoption Unit into an expanded Government Digital Service (GDS) within the Department for Science, Innovation and Technology (DSIT & GDS, 2025). This consolidation represents a significant shift from a federated approach towards centralised coordination of digital infrastructure, potentially addressing the longstanding fragmentation challenges.

## 4.5. Cross-case Analysis and Framework Validation

Our comparative examination reveals key patterns across ASA implementations while validating the framework's core principles. Three critical findings emerge from this analysis:

First, successful implementations maintain strategic coherence between objectives and execution, establish robust integration mechanisms between layers, and develop governance frameworks alongside technical capabilities. Estonia's X-Road and Singapore's Smart Nation demonstrate how effective coordination mechanisms enable systematic capability development.

Second, uneven development across layers creates significant implementation challenges. The UK's strong service design is constrained by fragmented infrastructure, while India's robust foundation layer outpaces governance mechanisms. Notably, the Intelligence and Process layers—where AI integration is most prominent—are generally less developed across most cases, with governance frameworks struggling to keep pace with technical innovation.

Third, the cases validate the ASA framework's systemic nature, where capabilities and constraints propagate across layers. Estonia's and Singapore's experiences demonstrate that integration mechanisms between layers are as important as the capabilities within each layer. The diversity of implementation pathways confirms the framework's flexibility—Estonia's infrastructure-first approach contrasts with the UK's service-driven transformation, each with distinct advantages and limitations.

These findings suggest refinements to enhance the framework's utility: explicitly addressing integration mechanisms between layers, providing contextual adaptation guidance for different resource constraints and legacy environments, and reconceptualising governance as an integral component embedded within each layer rather than a parallel concern.

# 5. Policy and Practice Implications

The ASA framework offers significant implications for policymakers and practitioners navigating AI-enabled government transformation. Effective governance requires tailored approaches that address the distinct challenges at each layer while maintaining cross-cutting coordination. Rather than treating governance as a separate concern, successful implementations embed governance mechanisms within technical architecture and operational processes.

Governance approaches must balance innovation with accountability across all layers. At the foundational level, this means addressing infrastructure sovereignty, security standards, and interoperability requirements. For data and intelligence systems, privacy-by-design principles must be combined with transparency mechanisms that reveal how insights are generated and applied. Algorithmic governance requires clear boundaries for automated decision-making, explainability standards, and human oversight mechanisms proportional to decision stakes. Service-level governance must ensure accessibility, inclusion, and meaningful public-private boundaries that preserve democratic accountability.

Strategic implementation requires careful attention to interdependencies between layers. organisations should begin by mapping current capabilities across all ASA layers to identify gaps and constraints, particularly where limitations in one layer may restrict development in others. This assessment informs prioritisation decisions about where to focus initial investments for maximum system-wide impact. While perfect equilibrium across layers is unrealistic, minimum viable capabilities must be maintained throughout the architecture to prevent critical bottlenecks.

Change management is crucial for effective transformation, encompassing both technical and organisational aspects. Stakeholder engagement must span technical experts, policymakers, operational staff, and end-users to ensure comprehensive input. Capability building should focus not only on technical skills but also on governance expertise that bridges technical implementation and policy objectives. Process redesign should explicitly address how AI integration changes workflows, responsibilities, and decision rights.

Performance measurement provides essential feedback for ongoing improvement. Beyond layer-specific metrics, organisations should develop system-level indicators that evaluate cross-layer integration, governance effectiveness, and public value creation. These should include operational efficiency gains, service quality improvements, and broader societal impacts including equity outcomes. Evaluation frameworks should establish clear baselines, incorporate independent assessment mechanisms, and report transparently to all stakeholders.

These implications demonstrate that successful AI-enabled transformation requires not just technological innovation but careful attention to governance, integration, and organisational change.

## 5.1. Policy Recommendations

Drawing on both the ASA framework and case study analysis, we offer several overarching policy recommendations:

1. **Establish Integrated Governance**: Develop governance frameworks that span all ASA layers rather than addressing them in isolation
2. **Invest in Foundational Capabilities**: Prioritise robust digital infrastructure and data governance as enablers for higher-layer innovations
3. **Build Public Sector Capability**: Develop specialised expertise within government rather than relying exclusively on private sector vendors
4. **Embrace Progressive Implementation**: Adopt staged approaches that build capabilities incrementally while maintaining strategic direction
5. **Balance Innovation and Inclusion**: Design implementations that explicitly address digital divides and ensure equitable access

6. **Maintain Democratic Oversight**: Ensure that AI-enabled transformation strengthens rather than undermines democratic accountability
7. **Foster Ecosystem Development**: Create frameworks that enable collaboration across public, private, and civil society sectors

These recommendations emphasise that successful AI-enabled government transformation requires not just technological innovation but careful attention to governance, inclusion, and democratic values.

# 6. Conclusions and Future Directions

This paper has introduced the Algorithmic State Architecture (ASA) as a novel framework for conceptualising and implementing AI-enabled government transformation. By framing Digital Public Infrastructure, Data-for-Policy, Algorithmic Government, and GovTech as interdependent layers of a coherent system, the ASA framework offers both analytical precision and practical guidance for policymakers and practitioners.

Our analysis demonstrates that effective AI implementation in government requires attention to the systemic relationships between layers, where capabilities and constraints propagate throughout the architecture. The case studies reveal both the potential of integrated approaches and the challenges that arise when development is uneven across layers. They highlight that technical implementation and governance frameworks must evolve together, with particular attention to maintaining democratic oversight as algorithmic capabilities advance.

The ASA framework makes several contributions to both theory and practice. Theoretically, it bridges previously disconnected domains of digital government research, offering a unified perspective on how technical and governance elements interact across what were previously treated as separate domains. Practically, it provides structured guidance for implementation planning, capability assessment, and governance design that reflects the complex reality of AI-enabled transformation, helping policymakers identify critical dependencies and potential bottlenecks in their digital transformation strategies.

Looking ahead, several promising directions for future research emerge. While our framework has been tested primarily through retrospective case analysis, longitudinal studies examining how ASA implementations evolve over time would enhance understanding of development pathways and adaptation mechanisms. Additionally, more granular analysis of the specific technical and organisational interfaces between layers would strengthen implementation guidance. Finally, exploration of how different political and institutional contexts shape ASA implementation would enhance the framework's applicability across diverse governance systems.

Future research should also explore the relationships between the core ASA elements and the broader ecosystem components. This includes examining how rapidly evolving external factors (particularly AI technology companies) influence government implementation of the four core elements, and developing multi-level governance approaches that account for the principles, governance, and enabling layers surrounding the core architecture. The framework's adaptability across varying institutional, social, and political contexts represents a particularly important area for further investigation.

As AI capabilities continue to advance, particularly with the rapid development of foundation models, the relationship between technical systems and democratic governance will only grow more consequential. The ASA framework offers a structured approach to navigating this complex terrain, ensuring that technological innovation enhances rather than undermines effective, accountable, and inclusive government.

**Credits:** Zeynep Engin: conceptualisation (lead); formal analysis (lead); writing - original draft (lead). Jon Corwcroft: methodology (supporting); writing - review and editing (equal). David Hand: review and editing (equal). Philip Treleaven: review and editing (equal).

**Acknowledgements:** The author(s) would like to thank the Data for Policy CIC Team, Emily Gardner, and Pinar Ozgen for their kind support during the drafting process. We also acknowledge our collaborator, Morine Amutorine from the Datasphere Initiative, for her valuable contributions.

**Declaration of interests:** None

**Funding:** This research did not receive any specific grant from funding agencies in the public, commercial, or not-for-profit sectors.

**Declaration of generative AI and AI-assisted technologies in the writing process:** During the preparation of this work, the author(s) used Claude 3.5 Sonnet and Claude 3.7 Sonnet models to help improve the presentation, refining language and structure for a diverse readership. After using these tools, the author(s) reviewed and edited all AI-assisted text and take(s) full responsibility for the content of the published article.